\documentclass[10pt,conference]{IEEEtran} 
\IEEEoverridecommandlockouts
 \usepackage{amsmath}
\usepackage{algorithmic}
\usepackage{graphicx}
\usepackage{textcomp}
\usepackage{enumitem}
\usepackage[table]{xcolor}
\usepackage{multirow}
\usepackage{algorithm}
\usepackage{listings}
\usepackage{url}
\usepackage{hyperref}
\usepackage{graphicx}
\usepackage{booktabs}
\usepackage{makecell}
\usepackage{multirow}
\usepackage{adjustbox}
\usepackage{arydshln}
\usepackage{threeparttable}
\usepackage{flafter} 
\def\BibTeX{{\rm B\kern-.05em{\sc i\kern-.025em b}\kern-.08em
    T\kern-.1667em\lower.7ex\hbox{E}\kern-.125emX}}
\begin{document}
\def\methodName#1{GapForge{#1}}
\newcommand{\mingxuan}[1]{{\color{red} #1 --Mingxuan}}
\setlength{\textfloatsep}{8pt plus 1pt minus 2pt}  
\setlength{\floatsep}{8pt plus 1pt minus 2pt}      
\setlength{\intextsep}{8pt plus 1pt minus 2pt}

\title{GapForge: Directed Compiler Fuzzing via Coverage-Gap Analysis}

\author{\IEEEauthorblockN{Anonymous}}

\author{\IEEEauthorblockN{1\textsuperscript{st} Mingxuan Zhu}
\IEEEauthorblockA{\textit{Key Lab of High Confidence Software Technologies}\\
\textit{(Peking University), Ministry of Education}\\
\textit{Peking University}\\
Beijing, China \\
zhumingxuan@stu.pku.edu.cn}
\and

\IEEEauthorblockN{2\textsuperscript{nd} Qingyuan Liang}
\IEEEauthorblockA{\textit{Key Lab of High Confidence Software Technologies}\\
\textit{(Peking University), Ministry of Education}\\
\textit{Peking University}\\
Beijing, China \\
liangqy@pku.edu.cn}
\and

\IEEEauthorblockN{3\textsuperscript{rd} Junjie Chen}
\IEEEauthorblockA{\textit{College of Intelligence and Computing} \\
\textit{Tianjin University}\\
Tianjin, China \\
junjiechen@tju.edu.cn}
\and

\IEEEauthorblockN{4\textsuperscript{th} Zhihong Xue}
\IEEEauthorblockA{\textit{Center Research Institute} \\
\textit{ZTE Corporation}\\
Shenzhen, China \\
13730826131@139.com}
\and

\IEEEauthorblockN{5\textsuperscript{th} Dan Hao\thanks{Dan Hao is the corresponding author.}}
\IEEEauthorblockA{\textit{Key Lab of High Confidence Software Technologies}\\
\textit{(Peking University), Ministry of Education}\\
\textit{Peking University}\\
Beijing, China \\
haodan@pku.edu.cn}
}

\maketitle

\begin{abstract}
Due to the scale and complexity of modern compiler codebases (e.g., GCC and LLVM), achieving comprehensive coverage across diverse code regions remains highly challenging. Most existing test generation techniques fail to exploit characteristics of the target code, producing test inputs (a.k.a. test programs) that exercise only a limited subset of all code. As a result, substantial regions of the compiler remain insufficiently tested, leading to persistent long-tail coverage gaps that are difficult to eliminate even across multiple releases. Although some white-box techniques have been proposed, they still achieve limited coverage when applied to large-scale compiler codebases.
To further improve compiler test coverage, particularly for exercising hard-to-reach edge regions, we present \methodName, a targeted LLM-based test generation technique that reasons about coverage gaps.
Unlike program-driven techniques that generate diverse inputs without modeling which compiler regions they exercise, and unlike whole-file summarization that yields only coarse guidance, \methodName\ treats coverage gaps as explicit region-level targets in three steps. First, it prioritizes promising files through coverage-driven scoring that favors large, under-covered files. Second, rather than summarizing a file as a whole, it pairs each uncovered line span with its enclosing covered context and performs path-difference analysis to infer fine-grained triggering requirements—both the program structures and the compilation options needed to reach the uncovered region. Third, it synthesizes prompts from these requirements together with prompts that previously failed on the same file, and uses the resulting coverage feedback to guide next-round selection.
To evaluate the performance of the proposed approach
\methodName, we conducted extensive experiments on GCC 14.3.0 and LLVM 19.1.0. The results show that \methodName\ significantly outperforms eight state-of-the-art techniques, demonstrating its effectiveness and generality for large-scale compiler testing. Within 72 hours, \methodName\ achieves 68.13\% and  69.11\% coverage on core compiler modules in GCC and LLVM, surpassing the state-of-the-art white-box technique WhiteFox by covering an additional 24,736 and 19,798 lines, respectively. Moreover, \methodName\ discovers 12 real-world compiler failures (5 in GCC and 7 in LLVM), including 8 crashes and 4 miscompilations, and each component of \methodName\ contributes to its performance. 
\end{abstract}


\section{Introduction}\label{sec:introduction}

Compilers are foundational system software that translate high-level languages into machine code. However, compiler defects can affect any program compiled, potentially leading to severe consequences such as crashes, silent miscompilations, and memory-safety violations~\cite{Survey}, making compiler correctness critical for software reliability.

\begin{figure*}[t!]
  \centering
  \includegraphics[width=0.76\textwidth]{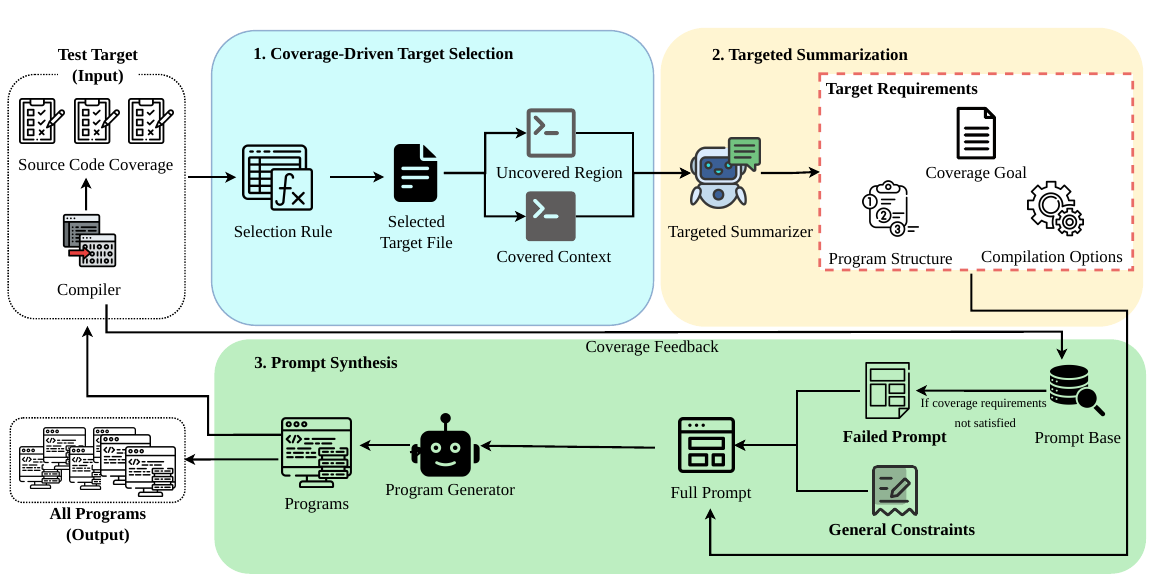}
  \caption{Workflow of \methodName.}
  \label{fig:workflow}
\end{figure*}

Consequently, many researchers have proposed techniques to generate effective test programs that expose latent defects~\cite{Csmith,CsmithEdge,YARPGen,LangFuzz,EMI,Creal,GrayC,Tzer}.\footnote{In this context, ``test programs'' refer to compiler inputs.}  Existing approaches fall into three categories: grammar-aided techniques~\cite{Csmith,CsmithEdge, YARPGen,LangFuzz} generate syntactically valid programs via language grammars; mutation-based techniques~\cite{EMI,Creal,GrayC,Tzer,Optimuzz} modify 
existing programs through semantics-preserving or semantics-altering transformations; and LLM-based techniques~\cite{Fuzz4All,WhiteFox,ni2025legofuzz} synthesize test programs using carefully designed prompts. These techniques have collectively uncovered numerous real-world compiler bugs.

However, despite their effectiveness in bug finding, existing techniques remain largely program-driven and lack compiler-wide, coverage-aware guidance. They mainly aim to generate diverse or bug-triggering inputs, but do not explicitly identify uncovered compiler files or code regions. Consequently, generated tests repeatedly exercise frequently triggered compilation paths (i.e., hotspot regions), while rarely executed code remains insufficiently tested and accumulates latent issues across multiple versions, making such defects increasingly difficult to identify and fix~\cite{CsmithEdge,bohme2020,regehr2016}.

More recently, the white-box technique WhiteFox~\cite{WhiteFox} leverages LLMs to summarize compiler optimization code and generate optimization-triggering tests. However, WhiteFox summarizes each source file as a whole and is primarily designed for optimization-related code, rather than systematically improving coverage across the entire compiler. As a result, it provides limited guidance for the large number of under-tested files and hard-to-reach code regions that lie outside optimization passes.

These limitations stem from two key challenges in compiler test generation: modern compilers contain a large number of source files with highly uneven coverage potential, making it essential to prioritize promising targets rather than test indiscriminately; moreover, reaching a specific uncovered region often requires not only the right program structure but also the right compilation configuration, which is difficult to infer from coarse, file-level analysis.

To address these limitations, we present \methodName, a targeted compiler testing technique that reasons about coverage gaps to guide test generation for large-scale compilers. Given a compiler codebase and its coverage information, \methodName\ outputs a set of test programs that systematically improve source-code coverage. 
The key idea is to treat coverage gaps as explicit targets and reason about how to reach them, through three steps: 
\methodName\ first \textbf{selects a promising target file} based on coverage feedback (i.e., assigns each source file a selection probability based on coverage ratio and code size), then \textbf{infers fine-grained triggering requirements} for the uncovered regions within each file, including the target file functionality and the required compilation options, and finally \textbf{synthesizes prompts} by integrating these requirements and prompts that previously failed to improve coverage on the same target file to generate effective test programs. By iterating this process, \methodName\ progressively exercises hard-to-reach code regions that existing techniques leave untested.

To evaluate \methodName, we conduct experiments on GCC~14.3.0~\cite{GCC} and LLVM~19.1.0~\cite{LLVM}. We compare \methodName\ against eight representative baselines, including Csmith~\cite{Csmith}, DST~\cite{DST}, Creal~\cite{Creal}, Fuzz4All~\cite{Fuzz4All}, LegoFuzz~\cite{ni2025legofuzz}, and the state-of-the-art white-box technique WhiteFox~\cite{WhiteFox}. Within 72 hours, \methodName\ achieves coverage of 68.13\% and 69.11\% in the core GCC and LLVM modules, and further improves the compilers own official test suites by 3,452 and 531 newly covered lines, outperforming all baselines. 
In contrast, the state-of-the-art white-box technique WhiteFox reaches only 64.62\%/65.02\% coverage, covering 24,736/19,798 fewer lines than \methodName{} and fails to improve the official test suite on GCC while adding only 55 lines on LLVM. The strongest baseline overall, LegoFuzz, reaches only 64.99\%/66.59\% coverage and improves the official test suites by merely 705/143 newly covered lines.
For the ablation analysis, we construct nine variants of \methodName\ by varying its components, and \methodName\ achieves better coverage results than all variants, demonstrating the impact of each component.

The contributions of this paper are summarized below.
\begin{itemize}[noitemsep, topsep=2pt]

  \item \textbf{An LLM-based compiler test generation technique, \methodName}, that targets coverage gaps by selecting promising coverage targets and inferring coverage-triggering requirements through combining execution coverage feedback with fine-grained analysis of uncovered code regions.
  \item \textbf{An extensive experiment} on the latest version of GCC and LLVM, which demonstrates the performance of \methodName\ as well as its components.
  \item \textbf{A reproducible package}, available in a public repository~\cite{repo}.
\end{itemize}
\section{Approach}\label{sec:approach}

Figure~\ref{fig:workflow} presents an overview of \methodName, a compiler testing technique to systematically improve the coverage of the source-code of large-scale compilers. The input to \methodName\ is a target compiler under test, given as its source codebase together with the line-coverage information of an initial run (i.e., from the compiler's existing test suite). The output is a set of test programs that, when compiled, exercise previously uncovered code regions and thereby improve overall coverage. 

In particular, \methodName\ operates over the whole codebase but processes it iteratively, selecting a single target file in each iteration and generating test programs aimed at the uncovered regions of that file.
Each iteration comprises three components: \textbf{(1) coverage-driven target selection}, which selects a promising source file and its uncovered code regions as the coverage target for each iteration; \textbf{(2) targeted summarization}, which analyzes the uncovered regions together with their covered context to infer fine-grained coverage-triggering requirements; and \textbf{(3) prompt synthesis}, which assembles a structured prompt from the inferred requirements and previously failed prompts to guide test-program generation.
We use the term \emph{targeted} to indicate that \methodName\ explicitly reasons about and steers generation toward specific uncovered code regions, in contrast to general-purpose fuzzers that generate diverse inputs without explicitly modeling which compiler paths they intend to exercise.

\subsection{Coverage-Driven Target Selection}\label{sec:selection}

At the start of each iteration, coverage-driven target selection chooses a single source file~$f$ from the codebase, along with its uncovered code regions, as the coverage target. This step is necessary because source files differ greatly in their coverage potential. Without principled selection, test generation may waste iterations on already well-covered files or on files whose uncovered paths are intrinsically hard to trigger, yielding little overall gain within a fixed time budget.

\smallskip
\noindent\textbf{Exploration score.}
\methodName\ assigns to each file $f$ in the codebase an exploration score $S_f$ that reflects its potential for coverage improvement:
\begin{equation}
    S_f = L_f \times (1 - C_f)^2
    \label{eq:score}
\end{equation}
where $L_f$ is the number of lines of code in file $f$ and $C_f \in [0,1]$ is the current line coverage ratio of file $f$. Both 
quantities are measured per file. The quadratic term $(1 - C_f)^2$ applies a non-linear transformation that amplifies differences among low-coverage files: a file at 10\% coverage receives a score roughly 81$\times$ higher than one at 90\% coverage with the same size.

\smallskip
\noindent\textbf{Selection weight and probability.}
To compare files against one another, \methodName\ normalizes each file's score across all files in the codebase into a selection weight:
\begin{equation}
    W_f = \frac{S_f}{\sum_{j=1}^{n} S_j}
    \label{eq:weight}
\end{equation}
where $n$ is the total number of source files in the codebase. While $W_f$ encodes relative importance, sampling directly from $W_f$ at each iteration causes files with the highest scores to dominate selection across iterations, leaving other potentially improvable files unvisited.

To prevent this over-concentration, \methodName\ applies a non-linear transformation to obtain the selection probability $P_f$:
\begin{equation}
    P_f = 1 - (1 - W_f)^k
    \label{eq:prob}
\end{equation}
$P_f$ can be interpreted as the probability that file $f$ would be selected at least once if $k$ independent draws were made using $W_f$.\footnote{Here $k$ is not the number of files selected per iteration-\methodName\ still draws exactly one-but a tunable parameter controlling how sharply the selection favors high-weight files.}
Crucially, this transformation \emph{compresses} the gap between 
high- and low-weight files: each additional draw yields diminishing returns for a high-weight file, while low-weight files receive a proportionally larger boost. \methodName\ normalizes $\{P_f\}$ into a final sampling distribution and draws one file per iteration. We set $k = 10$, for example, a file with $W_f = 0.08$ achieves $P_f = 56.56\%$.

After selecting file $f$, \methodName\ retrieves its uncovered code regions and partitions them into basic blocks, which serve as fine-grained coverage targets in the next step.

\begin{figure*}[t]
  \centering
  \includegraphics[width=0.96\textwidth]{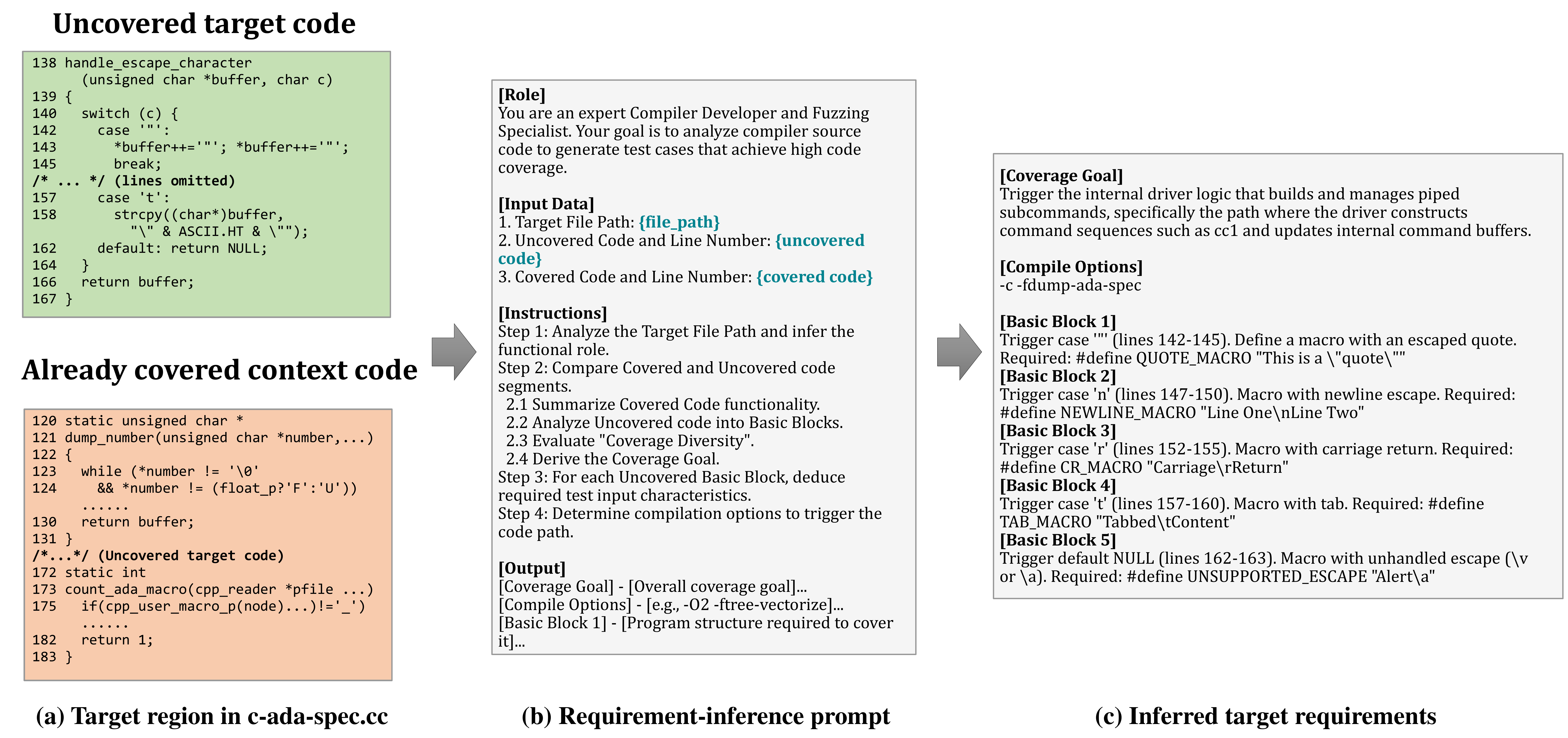}
  \caption{Targeted summarization for \texttt{c-ada-spec.cc}.}
  \label{fig:summarize}
\end{figure*}

To make this concrete, consider the GCC source file \texttt{c-ada-spec.cc} shown in Figure~\ref{fig:summarize} (a), which emits Ada specifications from C declarations. Among all files in the codebase, it receives a high exploration score: moderately sized yet low in coverage, so \methodName\ samples it as the target file and locates an uncovered region: the function \texttt{handle\_escape\_character} (lines 138--167), which remains unexercised by the existing test suite. We follow this example through the remaining components.

\subsection{Targeted Summarization}\label{sec:3-2}

Given the selected file and its uncovered regions, targeted summarization uses an LLM to infer \textit{Target Requirements}: a structured specification comprising a high-level coverage goal, recommended compilation options, and per-basic-block program-structure constraints needed to exercise the uncovered code. This component addresses two key challenges in improving compiler coverage: (1)~compiler files are large and redundant, making whole-file summarization too coarse-grained to pinpoint specific uncovered regions; and (2)~many regions are reachable only under specific compilation options and triggering program structures, which are difficult to identify without targeted analysis. To address these challenges, \methodName\ derives fine-grained coverage-oriented requirements that align with the target region and its compilation conditions.

\smallskip
\noindent\textbf{Covered context selection.}
Rather than processing the entire file, \methodName\ provides the LLM with each uncovered region together with its covered context. We work at the granularity of uncovered line spans: a coverage target is a contiguous run of uncovered lines, which may be only a few lines within an otherwise well-covered function (e.g., a single untaken branch). 
The covered context is obtained by scanning upward and downward from the uncovered span until a covered line is reached in each direction and taking the covered code that immediately encloses the span. This locally-adjacent code is the most relevant context, as it encodes the preconditions: variable states, branch conditions, and control-flow structure that execution must satisfy to reach the uncovered region.

In the prompt, the two are presented as separate code blocks, one labeled as the uncovered target and one as the covered context, each given with its original line numbers (as illustrated in Figure~\ref{fig:summarize} (a)). This lets the LLM perform path-difference analysis: reasoning about what input characteristics would steer control flow into the uncovered region rather than along the surrounding covered path.

\smallskip
\noindent\textbf{Requirement inference.}
\methodName\ prompts the LLM to perform a structured four-step analysis (shown in Figure~\ref{fig:summarize} (b)): (i)~infer the functional role of the target file in the compiler pipeline 
(e.g., front-end parsing, sanitizer instrumentation, or debug-info emission), providing global semantic grounding for subsequent reasoning;\footnote{This is the only step resembling prior whole-file summarization~\cite{WhiteFox}, yet it serves a different purpose: \methodName\ uses the functional role merely as coarse semantic context to ground the subsequent region-level analysis rather than as the end product driving generation.}(ii)~perform contrastive analysis between covered and uncovered paths to identify structural differences in control flow and data dependencies; (iii)~derive per-basic-block input requirements (e.g., specific data types, literal patterns, or control-flow constructs) needed to exercise each uncovered block; and (iv)~identify compilation options required to activate the target paths (e.g., \texttt{-fsanitize=address} for sanitizer-gated code), combining target-code semantics with path-difference analysis to derive option constraints.

Finally, \methodName\ produces the \textit{Target Requirements} $t$ (shown in Figure~\ref{fig:summarize} (c)): a structured specification including a whole-file coverage goal, block-specific program-structure requirements, and recommended compilation options, which guide subsequent prompt synthesis.

Returning to \texttt{c-ada-spec.cc}, Figure~\ref{fig:summarize} 
illustrates this step. The uncovered target \texttt{handle\_escape\_character} (lines 138--167) handles different escape characters through a \texttt{switch(c)} statement. \methodName\ selects the covered code in \texttt{dump\_number} (lines 120--131) and \texttt{count\_ada\_macro} (lines 172--184) as covered context. Through path-difference analysis, the LLM infers that each uncovered branch is reached by a macro with a specific escape sequence (e.g., \texttt{'\textbackslash n'}, \texttt{'\textbackslash t'}), and that the option \texttt{-fdump-ada-spec} must be enabled.

\begin{figure*}[t]
  \centering
  \includegraphics[width=0.96\textwidth]{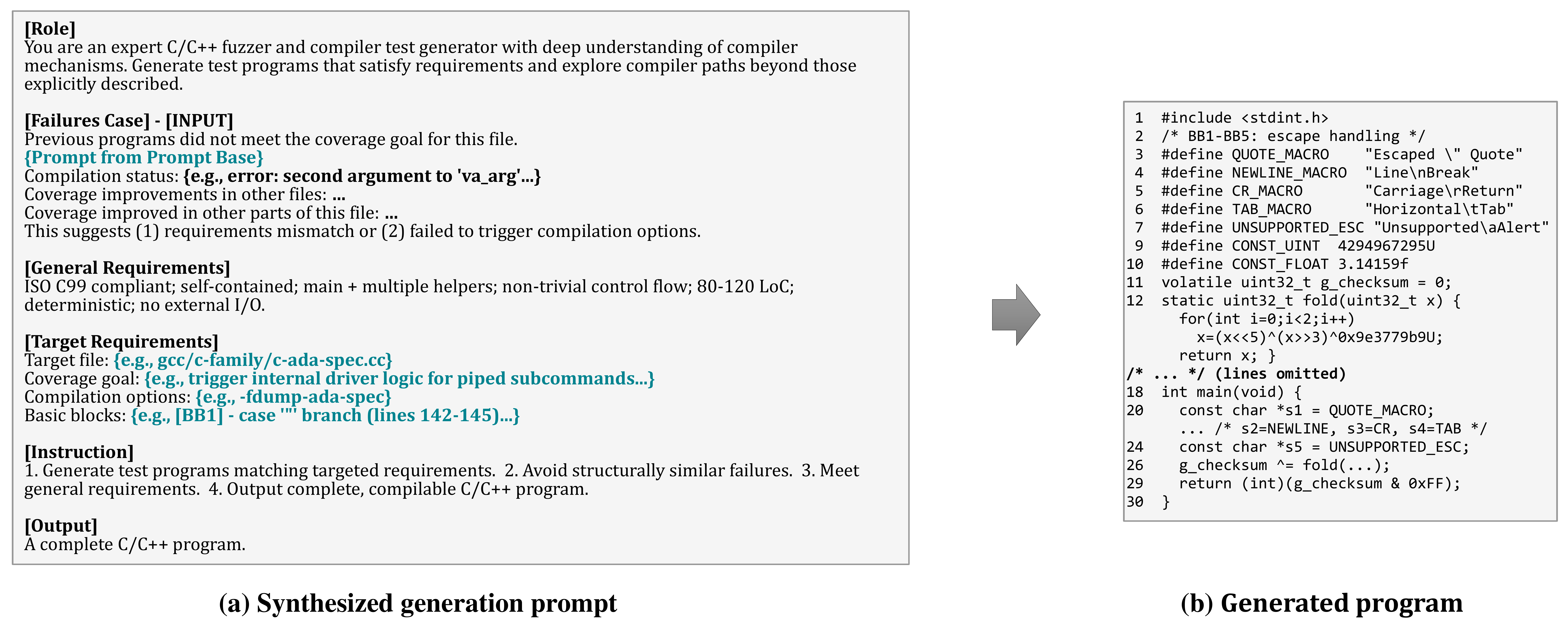}
  \caption{Prompt synthesis and program generation for \texttt{c-ada-spec.cc}.}
  \label{fig:generate}
\end{figure*}

\subsection{Requirement-Guided Prompt Synthesis with Failure Reflection}\label{sec:3-3}

Based on the \textit{Target Requirements} $t$ derived in Section~\ref{sec:3-2}, \methodName\ assembles a structured prompt from three elements: \textit{general program constraints} $r$, the \textit{Target Requirements} $t$, and \textit{previously failed prompts} $c$.

\smallskip
\noindent\textbf{General constraints.}
\methodName\ explicitly defines an expert role in the prompt to align the LLM's reasoning with compiler testing, encouraging coverage-oriented 
and compiler-specific outputs. The general constraints $r$ enforce that all generated programs include a \texttt{main} function, multiple helper functions, and non-trivial control flow (e.g., nested conditionals and loops). Program length is bounded within a moderate range to control the token budget while ensuring sufficient structural complexity to trigger deep compiler paths, according to findings that bug-triggering programs tend to be short yet structurally complex~\cite{Csmith}. All programs must be deterministic and free of external I/O dependencies to ensure reproducible coverage measurement.

\smallskip
\noindent\textbf{Target requirements.}
The \textit{Target Requirements} $t$ guide the prompt to focus on uncovered regions of the selected file, providing both high-level coverage objectives and fine-grained details including compilation options and basic-block-level triggering structures.

\smallskip
\noindent\textbf{Failure reflection.}
Targeted generation does not always exercise the intended uncovered 
regions. To mitigate this, \methodName\ maintains a \textit{per-file failure log} that records the prompts from earlier iterations of the current campaign that targeted the same file but failed to improve its coverage. 
A prompt is recorded as a failure when the generated program either 
(a)~fails to compile, or (b)~compiles and executes without covering any new target basic blocks. For each failure, \methodName\ records the prompt, the compilation error message (if any), and which regions were incidentally covered (if any).

Retrieval operates at \emph{file-level} granularity rather than a finer region-level one, since failures on the same file typically reflect a shared misunderstanding of its structure or required options that remains informative even as the target region shifts. When a file is re-selected, \methodName\ retrieves one failure case at random among those not yet used for this file and incorporates it with an instruction to avoid structurally similar programs. This keeps the prompt concise while progressively exposing the LLM to different ineffective strategies, forming an \textit{iterative refinement loop} that steers generation away from unproductive program structures.

\smallskip
Based on the integrated general constraints, target requirements, and failure cases, \methodName\ assembles the final prompt (shown in Figure~\ref{fig:generate} (a)) using a fixed instruction template, directing the LLM to generate complete and compilable C/C++ programs that match the target requirements and avoid previously failed structures. \methodName\ then compiles the generated programs with the specified options and collects coverage information to guide the next iteration.

For \texttt{c-ada-spec.cc}, Figure~\ref{fig:generate} shows the 
synthesized prompt and the generated program. In the current iteration, the failed prompt is appended to the new prompt 
with an instruction to avoid the same omission, leading the LLM to 
additionally define \texttt{NEWLINE\_MACRO}, \texttt{CR\_MACRO}, 
\texttt{TAB\_MACRO}, and \texttt{UNSUPPORTED\_ESC}, which together cover the remaining branches identified in the previous step.

\section{Experimental Setup}\label{sec:experiment_setup}

To evaluate the performance of \methodName, this experiment is designed to answer the following research questions:
\begin{itemize}
    \item \textbf{RQ1}: How effective is \methodName\ in improving compiler coverage?
    \item \textbf{RQ2}: How do different components affect the effectiveness of \methodName? 
    \item \textbf{RQ3}: How effective is \methodName\ in identifying compiler failures?
    \item \textbf{RQ4}: How does the choice of LLM affect \methodName's effectiveness?
\end{itemize}

\subsection{Target Compiler}\label{sec:4-1}
Following prior work, we use the most widely used C compilers (i.e., GCC 14.3.0 and LLVM 19.1.0) as our testing targets.

Modern compilers include many architecture-/platform-specific and auxiliary components that are not instrumented or produce no coverage information under our build configuration, making whole-compiler coverage noisy and less informative.\footnote{Excluded parts include target-specific back-end code for non-x86 architectures, build scripts, and auxiliary tooling outside the main compilation pipeline that cannot be reliably exercised by generated C/C++ programs.} We therefore exclude such components and retain the core front-end, middle-end, and back-end modules along the main compilation pipeline, whose coverage can be reliably influenced by generated tests. The retained modules are summarized in Table~\ref{files}.

For GCC, they span the front end, middle end, and back end: \texttt{gcc/c/} and \texttt{gcc/cp/} implement the C/C++ front end, 
\texttt{gcc/c-family/} provides shared infrastructure for C-like languages, \texttt{gcc/} covers IR construction, optimizations, and code generation, and \texttt{gcc/common/} handles shared utilities. For LLVM, they cover the core middle-end and back-end components: \texttt{IR} (IR definitions and utilities), \texttt{Analysis} (control-flow, dominance, aliasing, and loop analysis), \texttt{Transforms} (scalar/loop/ vectorization passes), and \texttt{CodeGen} (lowering IR to target machine code).

\begin{table}[t]
  \centering
  \caption{Target folders of the tested compilers.}
  \label{files}
  \footnotesize
  \setlength{\tabcolsep}{6pt}
  \renewcommand{\arraystretch}{1.15}
  \begin{tabular}{@{} l p{0.75\linewidth} @{}}
    \toprule
    \textbf{Compiler} & \textbf{Target file} \\
    \midrule
    GCC &
    \makecell[l]{gcc/*.c (533,176 LOC), gcc/cp/*.c (119,331 LOC),\\
                 gcc/c/*.c (30,310 LOC), gcc/c-family/*.c (21,907 LOC),\\
                 gcc/common/*.c (12 LOC)} \\
    LLVM &
    \makecell[l]{IR/*.c (57,869 LOC), Analysis/*.c (67,745 LOC),\\
     Transforms/*.c (250,007 LOC),\\
     CodeGen/*.c (108,431 LOC)}\\
    \bottomrule
  \end{tabular}
\end{table}

\subsection{Implementation}\label{sec:4-2}
We implement \methodName\ in Python and collect test coverage using Gcov-13~\cite{gcov}. For the compared techniques, we directly use their publicly available implementations whenever possible (i.e., Csmith, DST, Creal, GrayC, Fuzz4All, LegoFuzz and Optimuzz); otherwise, we re-implement the technique strictly following its paper description (i.e., WhiteFox).\footnote{WhiteFox originally targets optimization-related compiler files and does not define target selection for compiler components. So that, we add a neutral target-provision step: at each iteration, we uniformly sample one source file from the evaluation files and feed the corresponding code to WhiteFox, while keeping its subsequent steps unchanged.}

Our approach is general and thus agnostic to the choice of LLMs. In our implementation, we use GPT-4o for target summarization and StarCoder for test program generation following prior work~\cite{Fuzz4All,WhiteFox,ni2025legofuzz}. Importantly, this LLM configuration is consistent with that of the LLM-based baselines, enabling a fair comparison: Fuzz4All and WhiteFox use GPT-4o for prompt construction and StarCoder for test generation, while LegoFuzz uses GPT-4o to summarize mutation strategies. Thus, all of the compared LLM-based techniques are evaluated with the same underlying models for their corresponding LLM-driven components.

The experiment is performed on a workstation with an Intel Xeon Gold 6430 CPU (16 vCPUs), one NVIDIA RTX 4090 GPU with 24 GB of video memory, 120 GB of system memory, and Ubuntu 22.04.3 LTS operating system.

\subsection{Compared Techniques}\label{sec:4-3}

For RQ1 and RQ3, we compare \methodName\ with eight representative baselines: (1) \textbf{Csmith}~\cite{Csmith}, a grammar-based random program generator; (2) \textbf{DST}~\cite{DST}, a Csmith-based approach that adaptively adjusts configuration probabilities based on file-level coverage; (3) \textbf{GrayC}~\cite{GrayC}, a coverage-directed greybox fuzzer that mutates C programs under grammar and semantic constraints; (4) \textbf{Creal}~\cite{Creal}, a real-code-driven seed-augmentation generator that preserves semantics; (5) \textbf{Fuzz4All}~\cite{Fuzz4All}, an LLM-guided multi-objective fuzzing framework; (6) \textbf{WhiteFox}~\cite{WhiteFox}, a white-box LLM-guided generator that synthesizes optimization trigger tests; (7) \textbf{LegoFuzz}~\cite{ni2025legofuzz}, which uses LLM to construct and compose semantic components; and (8) \textbf{Optimuzz}~\cite{Optimuzz}, a directed greybox fuzzing framework that generates LLVM IRs to target specific compiler optimizations via translation validation.\footnote{Since Optimuzz generates LLVM IR rather than C/C++ source programs, it is only applicable to LLVM and is therefore excluded from the GCC comparison.} For a fair comparison, we assign a fixed time budget of 72 hours to all techniques for test generation.

Moreover, to fairly compare with the white-box technique WhiteFox (both directly leverage compiler source code), we integrate our coverage-driven file-level target selection into WhiteFox’s pipeline and construct a variant, denoted as $\mathrm{WhiteFox}_{Selection}$. This adaptation ensures a fair comparison by aligning the file-selection strategy across the two approaches.

For RQ2, we compare \methodName\ with ablated variants to investigate the contribution of each key component. Specifically, we conduct ablation studies by selectively removing each independent module and re-running the experiments for 24 hours. The variants are as follows:

\begin{itemize}
\item \textbf{Coverage-Driven target selection.} These variants examine how different target selection strategies affect the effectiveness of \methodName.
\begin{itemize}
\item $\mathrm{GapForge}_{NFS}$: Randomly selects coverage target.
\item $\mathrm{GapForge}_{LS}$: Linear scoring $S_f = L_f \times (1 - C_f)$ for target selection.
\item $\mathrm{GapForge}_{WS}$: Sampling directly from $W_f$ without converting to $P_f$.
\item $\mathrm{GapForge}_{5}$: Selection-probability transformation with $k = 5$.
\item $\mathrm{GapForge}_{15}$: Selection-probability transformation with $k = 15$.
\end{itemize}

\item \textbf{Target-context summarization.} These variants evaluate whether targeted summarization help \methodName\ generate more effective prompts.
\begin{itemize}
\item $\mathrm{GapForge}_{NS}$: No targeted summarization.
\item $\mathrm{GapForge}_{NCC}$: Summarization without covered context.
\item $\mathrm{GapForge}_{NO}$: No compilation option recommendation.
\end{itemize}

\item \textbf{Feedback-guided prompt refinement.} These variants study the impact of feedback information used to refine subsequent prompts.
\begin{itemize}
\item $\mathrm{GapForge}_{NFC}$: No previously failed prompt reflection.
\end{itemize}
\end{itemize}

For RQ4, we evaluate the generalization of \methodName\ across different LLMs. As described in Section~\ref{sec:4-2}, our default configuration 
uses GPT-4o for targeted summarization and StarCoder for test program generation. To assess sensitivity to the summarization model, we replace GPT-4o with DeepSeek-v3.2 and Qwen3-max, yielding two variants: $\mathrm{GapForge}_{ds}$ and $\mathrm{GapForge}_{qwen}$. To assess sensitivity to the generation model, we replace StarCoder with Qwen2.5-Coder-14B and Qwen2.5-Coder-32B, yielding two additional variants: $\mathrm{GapForge}_{qwen14}$ and $\mathrm{GapForge}_{qwen32}$. All variants are re-run under a 24-hour time budget.

\subsection{Measurement}\label{sec:4-4}
Following prior work~\cite{Fuzz4All,ni2025legofuzz}, we evaluate the coverage capability of compiler testing techniques using line coverage. Specifically, we use \texttt{gcov} to collect line coverage for all target source files (i.e., \textit{*.cc} and \textit{*.cpp}) within the target directories. In addition, we assess bug-finding capability by monitoring compiler crashes and performing differential testing under different optimization levels (i.e., \texttt{-O0}, \texttt{-O1}, \texttt{-O2}, and \texttt{-O3}). Any inconsistency in program output across optimization settings is treated as a potential compiler bug.

\section{Results}\label{sec:results_analysis}
\subsection{Overall Effectiveness (RQ1)}\label{sec:5-1}
\begin{table*}[!t]
\centering
\small
\setlength{\tabcolsep}{3pt}
\renewcommand{\arraystretch}{1.15}
\caption{Coverage results of the compared techniques.}
\begin{center}
{\footnotesize 
\textbf{Coverage from scratch}: \#Programs (generated test programs), $Cov_{lines}$ (covered lines), $Cov$(\%) (coverage ratio). \\
\textbf{Incremental improvement} over the original test suite: $\Delta Cov_{lines}$ (newly covered lines), $\Delta Cov$(\%) (ratio improvement).\\
\textbf{Tokens}: LLM token consumption, ``-'' for 
non-LLM-based techniques.}
\end{center}
\resizebox{\linewidth}{!}{
\begin{tabular}{l || c:cc:cc:c | c:cc:cc:c| c:cc:cc:c}
\toprule
\multirow{2}{*}{} 
  & \multicolumn{6}{c}{24 hours} \vline
  & \multicolumn{6}{c}{48 hours} \vline
  & \multicolumn{6}{c}{72 hours} \\
\cmidrule(lr){2-7}\cmidrule(lr){8-13}\cmidrule(lr){14-19}
  & \#Programs & $Cov_{lines}$ & $Cov$ (\%) & $\Delta Cov_{\text{lines}}$ & $\Delta Cov (\%)$ & Tokens
  & \#Programs & $Cov_{lines}$ & $Cov$ (\%) & $\Delta Cov_{\text{lines}}$ & $\Delta Cov (\%)$ & Tokens
  & \#Programs & $Cov_{lines}$ & $Cov$ (\%) & $\Delta Cov_{\text{lines}}$ & $\Delta Cov (\%)$ & Tokens \\
\midrule
\multicolumn{19}{c}{\textbf{GCC}} \\
\midrule
Csmith   & 31,106 & 359,979 & 51.08\% & +0 & +0 & - & 60,673 & 359,979 & 51.08\% & +0 & +0 & - & 92,535 & 359,979 & 51.08\% & +0 & +0 & - \\
DST      & 28,934 & 360,049 & 51.09\% & +0 & +0 & - & 57,015 & 360,049 & 51.09\% & +0 & +0 & - & 84,690 & 360,049 & 51.09\% & +0 & +0 & - \\
Creal    & 22,874 & 360,472 & 51.15\% & +0 & +0 & - & 45,488 & 360,754 & 51.19\% & +0 & +0 & - & 66,073 & 361,036 & 51.23\% & +0 & +0 & - \\
GrayC    & 21,965 & 386,124 & 54.79\% & +0 & +0 & - & 43,765 & 390,493 & 55.41\% & +93 & +0.08\% & - & 64,498 & 399,020 & 56.62\% & +164 & +0.13\% & - \\
Fuzz4All & 18,243 & 427,281 & 60.63\% & +0 & +0 & \textbf{186,392} & 35,791 & 438,416 & 62.21\% & +0 & +0 & \textbf{186,392} & 55,161 & 442,010 & 62.72\% & +0 & +0 & \textbf{186,392} \\
WhiteFox & 16,184 & 437,359 & 62.06\% & +0 & +0 & 799,436 & 32,335 & 448,141 & 63.59\% & +0 & +0 & 1,547,296 & 48,617 & 455,400 & 64.62\% & +0 & +0 & 2,304,896 \\
LegoFuzz & 20,286 & 450,819 & 63.97\% & +705 & +0.59\% & 3,628,548 & 41,694 & 452,581 & 64.22\% & +705 & +0.59\% & 3,628,548 & 61,374 & 458,007 & 64.99\% & +705 & +0.59\% & 3,628,548 \\
Optimuzz & - & - & - & - & - & - & - & - & - & - & - & - & - & - & - & - & - & - \\
GapForge & 14,571 & \textbf{474,780} & \textbf{67.37\%} & \textbf{+2,043} & \textbf{+1.71\%} & 337,149 & 27,535 & \textbf{476,753} & \textbf{67.65\%} & \textbf{+2,556} & \textbf{+2.14\%} & 644,368 & 42,288 & \textbf{480,136} & \textbf{68.13\%} & \textbf{+3,452} & \textbf{+2.89\%} & 976,364 \\
\midrule
\multicolumn{19}{c}{\textbf{LLVM}} \\
\midrule
Csmith   & 31,106 & 286,365 & 59.16\% & +0 & +0 & - & 59,673 & 286,365 & 59.16\% & +0 & +0 & - & 92,535 & 286,365 & 59.16\% & +0 & +0 & - \\
DST      & 27,159 & 289,120 & 59.73\% & +0 & +0 & - & 54,491 & 289,120 & 59.73\% & +0 & +0 & - & 81,124 & 289,120 & 59.73\% & +0 & +0 & - \\
Creal    & 22,874 & 291,738 & 60.27\% & +0 & +0 & - & 45,488 & 292,803 & 60.49\% & +0 & +0 & - & 66,073 & 292,803 & 60.49\% & +0 & +0 & - \\
GrayC    & 20,775 & 292,028 & 60.33\% & +83 & +0.18\% & - & 42,710 & 294,641 & 60.87\% & +102 & +0.22\% & - & 63,287 & 295,464 & 61.04\% & +102 & +0.22\% & - \\
Fuzz4All & 19,088 & 300,741 & 62.13\% & +0 & +0 & \textbf{164,269} & 38,629 & 300,741 & 62.13\% & +0 & +0 & \textbf{164,269} & 56,194 & 304,275 & 62.86\% & +0 & +0 & \textbf{164,269} \\
WhiteFox & 17,289 & 311,051 & 64.26\% & +0 & +0 & 379,513 & 35,258 & 313,230 & 64.71\% & +0 & +0 & 761,138 & 52,594 & 314,730 & 65.02\% & +55 & +0.12\% & 1,070,226 \\
LegoFuzz & 20,286 & 318,215 & 65.74\% & +96 & +0.21\% & 3,628,548 & 41,694 & 320,345 & 66.18\% & +96 & +0.21\% & 3,628,548 & 61,374 & 322,187 & 66.59\% & +143 & +0.32\% & 3,628,548 \\
Optimuzz & 21,239 & 317,440 & 65.58\% & +173 & +0.39\% & - & 43,259 & 324,217 & 66.98\% & +209 & +0.47\% & - & 63,615 &  325,330 & 67.21\% & +285 & +0.62\% & - \\
GapForge & 14,186 & \textbf{329,591} & \textbf{68.09\%} & \textbf{+389} & \textbf{+0.85\%} & 326,417 & 28,235 & \textbf{332,011} & \textbf{68.59\%} & \textbf{+466} & \textbf{+1.02\%} & 638,075 & 41,255 & \textbf{334,528} & \textbf{69.11\%} & \textbf{+531} & \textbf{+1.16\%} & 944,817 \\
\bottomrule
 \end{tabular}}

\label{t5-1}
\end{table*}

To answer RQ1, we compare the coverage achieved by different techniques under three time budgets (24, 48, and 72 hours) from two perspectives: 
coverage from scratch (\#Programs --- the number of generated test programs, $Cov_{lines}$ --- covered lines, $Cov$(\%) --- coverage ratio) and incremental improvement over the original compiler test suites ($\Delta Cov_{\text{lines}}$ --- covered-line improvement, $\Delta Cov$(\%) --- coverage-ratio improvement). Both perspectives are computed from the same set of generated test programs: coverage from scratch measures what these programs cover alone, while incremental improvement measures the additional lines they cover 
on top of the original test suite. The two differ only in the baseline, not in the programs evaluated.
For LLM-based techniques, we additionally report token consumption (Tokens) to characterize computational cost.\footnote{Non-LLM-based techniques (e.g., Csmith, GrayC) do not invoke LLM APIs and thus incur no token cost, denoted as ``-'' in the 
Tokens column.} We further analyze hourly coverage evolution during the first 24 hours to characterize coverage growth over time. We also perform an in-depth comparison with WhiteFox, and conduct case studies to explain why \methodName\ covers regions missed by other techniques.

Table~\ref{t5-1} summarizes the results on GCC and LLVM. For each budget, the best result is highlighted in bold. Under the 72-hour budget, \methodName\ achieves 68.13\% (480,136 lines) and 69.11\% (334,528 lines) coverage on GCC and LLVM, improving beyond the original test 
suites by 3,452 lines (+2.89\%) and 531 lines (+1.16\%), respectively.
Moreover, \methodName\ consistently increases the coverage of both compilers as the time budget increases from 24 to 72 hours.

\subsubsection{Coverage from scratch Analysis}\label{sec:5-1-1}

\begin{figure*}[t]
    \centering
    \begin{minipage}{0.48\linewidth}
        \centering
        \includegraphics[width=\linewidth]{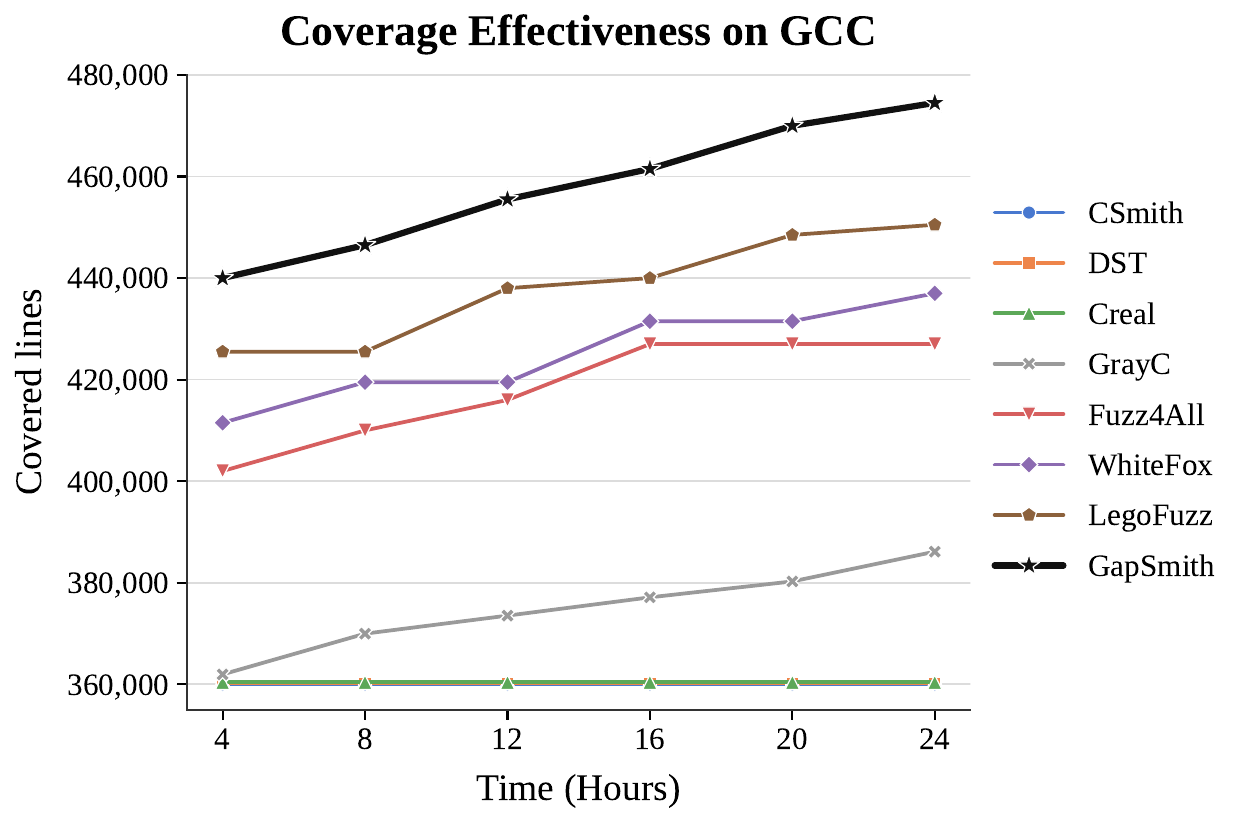}
        \vspace{2pt}
        \small (a) GCC results
    \end{minipage}
    \hfill
    \begin{minipage}{0.48\linewidth}
        \centering
        \includegraphics[width=\linewidth]{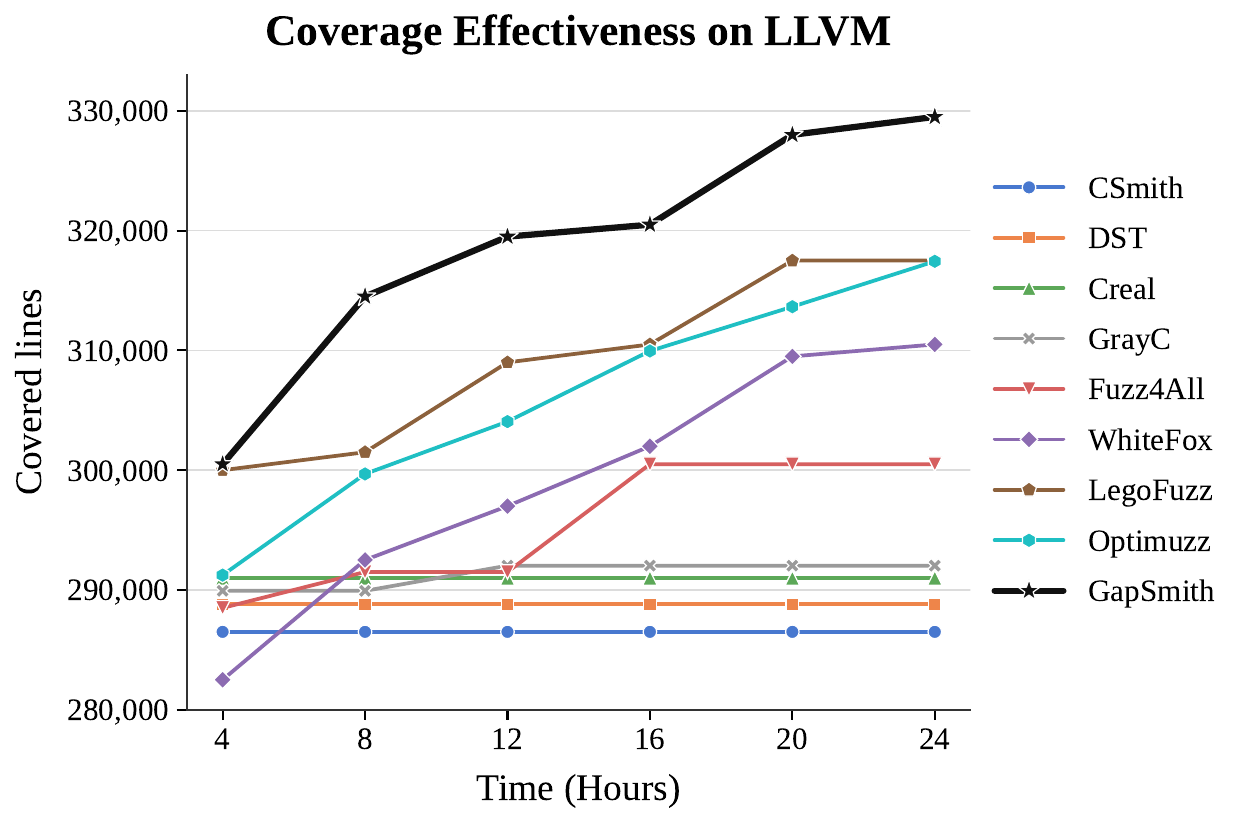}
        \vspace{2pt}
        \small (b) LLVM results
    \end{minipage}
    \caption{Coverage results on LLVM and GCC in 24 hours.}
    \label{coverfig}
\end{figure*}

As shown in Table~\ref{t5-1}, \methodName\ consistently achieves the best coverage despite generating the fewest test programs. Under the 72-hour budget, it covers 22,129/12,341 more lines than the strongest baseline LegoFuzz, and 120,157/48,163 more than the weakest baseline Csmith, on GCC/LLVM. Figure~\ref{coverfig} further shows that \methodName\ steadily widens this gap over time, indicating a superior coverage efficiency rather than a single advantage.

\subsubsection{Incremental Improvement Analysis}\label{sec:5-1-2}
Compilers already ship with massive test suites (e.g., GCC 14.3.0 includes 399,007 tests) that yield high baseline coverage (82.93\% on GCC and 90.55\% on LLVM core code), leaving little room for further improvement. 
We therefore measure the lines newly covered beyond these original suites. As shown in Table~\ref{t5-1}, \methodName\ still adds 3,452 and 531 new lines on GCC and LLVM under the 72-hour budget, whereas Csmith, DST, Creal, and Fuzz4All add none, and the best baseline LegoFuzz adds only 705 and 143. This shows that \methodName\ effectively reaches hard-to-cover long-tail regions that other techniques cannot.

To examine these long-tail regions directly, we identify \textit{hard-to-cover} files: those whose \textbf{average coverage across all techniques remains below 30\%}, representing systematic blind spots.
This yields $N_{gcc} = 152$ and $N_{llvm} = 123$ such files in GCC and LLVM, respectively.\footnote{Examples include \texttt{c-ada-spec.cc} and \texttt{module.cc} in GCC; \texttt{SelectionDAGPrinter.cpp} and \texttt{RegionPrinter.cpp} in LLVM.}
On these files, Csmith reaches only 8.38\%/10.55\% and even the strongest baseline WhiteFox only 27.43\%/29.65\% on GCC/LLVM, while \methodName\ reaches 36.29\%/39.75\%, confirming that explicit coverage-gap reasoning allows \methodName\ to exercise regions that resist all existing techniques.

Beyond coverage, \methodName\ is also token-efficient. LegoFuzz and Fuzz4All call the LLM only once upfront --- incurring a heavy 3.63M one-time cost for LegoFuzz versus a lightweight 186K for Fuzz4All --- while WhiteFox and \methodName\ query the LLM throughout. Owing to its concise input 
context, \methodName\ uses 976K tokens on GCC under 72 hours, less than half of WhiteFox's 2.30M, while achieving substantially higher coverage. In short, \methodName\ delivers stronger results at lower cost.

\subsubsection{Comparison with WhiteFox}\label{sec:5-1-3}

WhiteFox is the closest white-box technique to \methodName\, as both leverage compiler source code. We first ask whether \methodName{}'s coverage-driven target selection can benefit WhiteFox itself by integrating it into WhiteFox's pipeline, yielding $\mathrm{WhiteFox}_{Selection}$. 
While the original WhiteFox fails to improve over the existing test suites on either compiler, $\mathrm{WhiteFox}_{Selection}$ gains 979 and 178 lines on GCC and LLVM, showing that target selection generalizes beyond \methodName. Yet it still trails \methodName\ (63.84\%/66.02\% vs. 67.37\%/68.09\%), indicating that \methodName's region-level summarization and failure-aware prompt synthesis contribute substantially as well.

Since WhiteFox specifically targets compiler optimizations, we further compare on optimization code under a per-target setup: unlike the evaluation above, which measures overall coverage over the full file set under a fixed \emph{time} budget, we randomly select 50 LLVM optimization-specific files and generate a fixed 100 test programs per file for each technique, counting how many targets are exercised. Even on WhiteFox's home ground, \methodName{} covers 19 of the 50 targets versus WhiteFox's 11, confirming its stronger ability to reach hard-to-cover optimization regions.

\subsubsection{Case Study}\label{sec:5-1-4}

To understand why \methodName\ covers regions missed by others, we inspect 
representative tests targeting GCC's \textit{dwarf2out.cc}. This file implements DWARF debug-information emission, a core component invoked whenever programs are compiled with debugging enabled; with over 14,000 lines, it is one of the largest and most complex files in GCC, making it a representative rather than contrived target.
\methodName\ infers that the input should include bit-level builtins such as  \textit{\_\_builtin\_clz} (the target code constructs DWARF location expressions for CLZ-related values), and recommends \textit{-g} and \textit{-gdwarf-4} to force DWARF generation, as well as \textit{-fvar-tracking} and \textit{-fvar-tracking-assignments} to enhance variable tracking and trigger richer location expressions.

In contrast, WhiteFox only provides coarse file-level summaries and fails to derive these concrete triggering conditions, yielding little coverage improvement on the same file.

\subsection{Ablation Study (RQ2)}\label{sec:5-2}
To answer RQ2, we construct 9 variants by removing each component individually, evaluate them on GCC under a 24-hour budget, and report both coverage results and token consumption. Table~\ref{ablation} summarizes the results, with the best highlighted in bold.

\subsubsection{Impact of Target File Selection}
We study the target selection component through five variants: random selection ($\mathrm{GapForge}_{NFS}$), linear scoring ($\mathrm{GapForge}_{LS}$), direct weight sampling without probability transformation ($\mathrm{GapForge}_{WS}$), and two parameter settings $k=5$ ($\mathrm{GapForge}_{5}$) and $k=15$ ($\mathrm{GapForge}_{15}$). 
As shown in Table~\ref{ablation}, $\mathrm{GapForge}_{NFS}$ degrades the most (25,580 fewer covered lines and 1,733 fewer newly covered lines), as random selection repeatedly wastes iterations on hard-to-cover targets (e.g., \texttt{tree-vect-stmts.cc}) that yield no improvement. The other variants also fall short of \methodName{}: linear scoring ($\mathrm{GapForge}_{LS}$)) fails to differentiate low-coverage files, direct weight sampling ($\mathrm{GapForge}_{WS}$) lets high-scoring files dominate, $k=5$ over-concentrates on few files, and $k=15$ over-disperses effort. The default $k=10$ thus best balances exploration and exploitation.

\subsubsection{Impact of Targeted Summarization}
To investigate the contribution of targeted summarization, we construct three variants: $\mathrm{GapForge}_{NS}$, which directly uses uncovered code with its covered context as the \textit{Target Requirements}, $\mathrm{GapForge}_{NCC}$, which 
summarizes only the uncovered regions without their covered context, and $\mathrm{GapForge}_{NO}$, which removes both option recommendations during targeted summarization and the corresponding structural requirements in prompt synthesis.

Although $\mathrm{GapForge}_{NS}$ saves tokens and generates more programs, coverage drops by 16,730 lines, since summarization provides critical guidance on program structure, functional context, and compilation options.

In particular, $\mathrm{GapForge}_{NCC}$ reduces the coverage by 8,540 lines. The covered context supplies concrete control-flow and data-flow cues (e.g., branch conditions and variable usage) around uncovered blocks, which the model needs to infer program structures that trigger the target regions.

$\mathrm{GapForge}_{NO}$ suffers the largest drop among all variants (19,480 fewer covered lines), confirming that compilation options are critical for reaching option-gated paths: without the recommended \texttt{-fsanitize} flag, for instance, the target block in \texttt{asan.cc} stays unreachable regardless of program structure.



\subsubsection{Impact of Failure Case Reflection}
To investigate the contribution of failure case reflection, we construct $\mathrm{GapForge}_{NFC}$, which removes previously failed prompts from prompt synthesis. 
Its drop is modest (2,780 fewer lines) because target selection seldom revisits the same file within 24 hours, limiting how often failure cases accumulate: only 6 informative cases arise in 24 hours, but 22 in 72 hours. This suggests that failure reflection grows more beneficial under longer budgets or when generation focuses intensively on specific files.

\begin{table}[t]
\centering
\small
\setlength{\tabcolsep}{4pt}
\renewcommand{\arraystretch}{0.95}
\caption{Impact of each component of GapForge.}
\label{ablation}
\resizebox{\columnwidth}{!}{%
\begin{tabular}{l||c:cc:cc:c}
\toprule
\textbf{Technique} & \#Programs & $Cov_{lines}$ & $Cov$ (\%) & $\Delta Cov_{\text{lines}}$ & $\Delta Cov (\%)$ & \textbf{Token} \\
\midrule
\multicolumn{7}{c}{\textbf{GCC}} \\
\midrule
$\mathrm{GapForge}_{NFS}$ & 15,006 & 449,200 & 63.72\% & 310   & +0.26\% & 346,894 \\
$\mathrm{GapForge}_{LS}$  & 14,285 & 471,398 & 66.89\% & 1,266 & +1.06\% & 316,285 \\
$\mathrm{GapForge}_{WS}$  & 14,559 & 464,703 & 65.94\% & 1,207 & +1.01\% & 325,693 \\
$\mathrm{GapForge}_{5}$   & 14,631 & 472,314 & 67.02\% & 1,613 & +1.35\% & 336,794 \\
$\mathrm{GapForge}_{15}$  & 14,172 & 473,935 & 67.25\% & 1,816 & +1.52\% & 308,112 \\
\midrule
$\mathrm{GapForge}_{NS}$  & 17,268 & 458,050 & 64.99\% & 1,266 & +1.06\% & \textbf{0} \\
$\mathrm{GapForge}_{NCC}$ & 15,038 & 466,240 & 66.27\% & 1,827 & +1.53\% & 278,924 \\
$\mathrm{GapForge}_{NO}$  & 14,791 & 455,300 & 64.56\% & 919   & +0.77\% & 331,098 \\
\midrule
$\mathrm{GapForge}_{NFC}$ & 15,134 & 472,000 & 67.11\% & 1,825 & +1.53\% & 357,063 \\
\midrule
GapForge & 14,571 & \textbf{474,780} & \textbf{67.37\%} & \textbf{2,043} & \textbf{+1.71\%} & 337,149 \\
\bottomrule
\end{tabular}
}
\end{table}

\subsection{Failure Study (RQ3)}\label{sec:5-3}
To investigate the failure-finding effectiveness of different techniques, we perform differential testing on the target compilers using the test programs generated within a 24-hour budget, by compiling each program with different optimization levels (i.e., -O0, -O1, -O2, and -O3) and checking for inconsistent behaviors. We first calculate the total number of failures found by each technique as well as the number of unique failures. We then analyze the types of failures discovered by \methodName. Finally, we conduct a case study on representative failure-triggering programs generated by \methodName.

Table~\ref{failure_status} summarizes the number of failures reported by all compared techniques. Overall, on GCC 14.3.0 and LLVM 19.1.0, \methodName\ reports 5 and 7 failures, respectively. Compared with other LLM-based techniques (e.g., Fuzz4All and WhiteFox), \methodName\ reports more failures. Although LegoFuzz reports more failures, it relies on expensive offline database construction and iterative synthesis. In contrast, \methodName\ achieves a comparable failure-finding performance with much lower resource cost.

We categorize the reported failures into two types: (1) \emph{Crash}, where the compiler fails due to assertion/runtime errors, and (2) \emph{Miscompilation}, where the compiler silently generates incorrect code, which is particularly critical~\cite{Survey}. \methodName\ discovers 8 crash failures (4 in GCC and 4 in LLVM) and 4 miscompilation failures (1 in GCC and 3 in LLVM), indicating that its generated programs can trigger subtle and hard-to-detect compiler defects. Overall, these results show that although \methodName\ is designed for coverage improvement, it can also effectively expose real-world compiler defects.

\begin{table}[t]
\centering
\footnotesize
\caption{Detected failures.}
\label{failure_status}
\scalebox{0.85}{%
\begin{tabular}{l||ccc}
\toprule
\textbf{Technique} & \textbf{GCC} & \textbf{LLVM} & \textbf{All}\\
\midrule
Csmith   & 0 & 0 &  0 \\
DST      & 0 & 0 &  0 \\
Creal    & 2 & 4 &  6 \\
GrayC    & 1 & 2 &  3 \\
Fuzz4All & 1 & 1 &  2 \\
WhiteFox & 1 & 3 &  4 \\
LegoFuzz & 6 & 9 & 15 \\
Optimuzz & - & 6 &  6 \\
GapForge & 5 & 7 & 12 \\
\bottomrule
\end{tabular}%
}
\end{table}

To illustrate \emph{what kinds} of defects \methodName\ can expose, we 
present two representative cases it discovered. Both programs are reduced from the originals for readability.

\noindent\textbf{Crash in GCC.} While targeting \textit{reload.cc}, \methodName\ generated a program combining mutually exclusive specifiers (\texttt{\_Thread\_local}, \texttt{\_Noreturn}, and \texttt{register}) in a single declaration, triggering an internal compiler error.

\noindent\textbf{Miscompilation in LLVM.} While targeting \textit{ConstraintElimination.cpp}, \methodName\ generated a program 
with tightly coupled integer constraints across nested branches (e.g., 
chaining \texttt{(x < c1) \&\& (x > c2)} with bit-masking and conditional 
updates such as \texttt{x = (x \& m) + k}), which exercises LLVM's 
constraint-propagation logic and leads to a miscompilation.

\subsection{Effect of Different LLMs (RQ4)}\label{sec:5-5}

\begin{table}[t]
\centering
\caption{Impact of the base LLM of GapForge.}
\label{model}
\small
\setlength{\tabcolsep}{3pt}
\renewcommand{\arraystretch}{1.0}
\scalebox{0.8}{%
\begin{tabular}{l||c:cc:cc:c}
\toprule
\textbf{Technique} & \#Programs & $Cov_{lines}$ & $Cov$ (\%) & $\Delta Cov_{\text{lines}}$ & $\Delta Cov (\%)$ & \textbf{Token} \\
\midrule
GapForge & 14,571 & 474,780 & 67.37\% & 2,043 & +1.71\% & 337,149 \\
$\mathrm{GapForge}_{DS}$ & 16,014 & 465,394 & 66.08\% & 1,112 & +0.93\% & 379,028 \\
$\mathrm{GapForge}_{Qwen}$ & 15,848 & 466,816 & 66.24\% & 1,541 & +1.29\% & 364,667\\
$\mathrm{GapForge}_{qwen14}$ & 14,328 & 475,978 & 67.54\% & 2,138 & +1.79\% & 336,279\\
$\mathrm{GapForge}_{qwen32}$ & 14,267 & \textbf{478,868} & \textbf{67.95\%} & \textbf{2,425} & \textbf{+2.03\%} & \textbf{328,541}\\
\bottomrule
\end{tabular}
}
\end{table}

To answer RQ4, we instantiate \methodName\ with different LLMs under a 24-hour budget on GCC. To assess sensitivity to the summarization model, we replace GPT-4o with DeepSeek-v3.2 and Qwen3-max, yielding $\mathrm{GapForge}_{ds}$ and $\mathrm{GapForge}_{qwen}$. To assess 
sensitivity to the generation model, we replace StarCoder with Qwen2.5-Coder-14B and Qwen2.5-Coder-32B, yielding $\mathrm{GapForge}_{qwen14}$ and $\mathrm{GapForge}_{qwen32}$. We 
select these as representative strong models from different families; evaluating more LLMs is limited by API cost and runtime overhead.

Table~\ref{model} presents the results. For the summarization model, although $\mathrm{GapForge}_{ds}$ generates more test programs, it 
covers 9,385 fewer lines and achieves 931 fewer newly covered lines than \methodName\, suggesting that the quality of summarization has a notable 
impact on coverage effectiveness. For the generation model, $\mathrm{GapForge}_{qwen32}$ achieves competitive coverage compared with 
\methodName, demonstrating that a stronger code generation model can partially compensate for the difference in generation capability. 
Nevertheless, all variants consistently outperform the baseline techniques, demonstrating that \methodName's effectiveness is not overly sensitive to the choice of LLM.

\section{Threats to Validity}\label{sec:threats}
\textbf{The internal threat} arises mainly from the implementation of compiler test generation approaches. To reduce this threat, we utilize the reproducible packages of the compared approaches and strictly re-implement WhiteFox based on its corresponding publication. 

\textbf{The external threat} comes mainly from the target compilers and platforms. To reduce this threat, we use two popular compilers (i.e., GCC and LLVM), and most commonly the x86 platform, as in previous work~\cite{Fuzz4All,WhiteFox,Csmith,Creal,zhu2024compiler,zhu2025compiler}. 

\textbf{The construct threat} mainly comes from the evaluation metrics and measurement procedures. 
Following prior work~\cite{Fuzz4All,WhiteFox,ni2025legofuzz}, we use line coverage as the primary coverage metric and collect coverage using \texttt{gcov}. Since our approach is LLM-guided, we also report token consumption as a cost metric, although token usage may vary across APIs and implementations. For bug-finding evaluation, we monitor compiler crashes and perform differential testing under different optimization levels~\cite{Fuzz4All,WhiteFox,ni2025legofuzz}. 
Additionally, we acknowledge a potential hardware bias: traditional fuzzers such as Csmith, GrayC are optimized for CPUs, whereas LLM-based techniques rely on GPU acceleration. 
Since all experiments share the same hardware configuration, this difference may affect the number of test programs generated per unit time.
\section{Related Work}\label{sec:related}
Compiler testing is a fundamental problem in software testing, and extensive research has been conducted on test oracle construction and test generation. This paper focuses on compiler test generation. In this section, we first review existing compiler test generation techniques. Since our approach leverages LLMs, we further discuss representative LLM-based test generation methods for other software targets.

\subsection{Compiler Test Generation}\label{sec:8-1}

Compiler test generation techniques can be broadly categorized into grammar-aided, mutation-based, and LLM-based approaches~\cite{ni2025legofuzz}.

Grammar-aided techniques~\cite{Csmith,CsmithEdge,DST,YARPGen} generate syntactically valid programs by following language production rules and enforcing semantic constraints. Representative tools include Csmith~\cite{Csmith}, which constructs semantically complete C programs by assigning probabilities to different syntactic structures, and 
YARPGen~\cite{YARPGen}, which uses generation policies to target scalar 
and loop optimizations.

Mutation-based techniques~\cite{stochastic,EMI,Creal,GrayC,Optimuzz} are also effective for discovering compiler bugs. EMI~\cite{EMI} mutates seed programs while preserving input-output behaviors, exposing bugs in GCC and LLVM. Creal~\cite{Creal} injects real-world functions into seed programs in a semantics-preserving manner. GrayC~\cite{GrayC} incorporates coverage 
feedback and mutates programs under grammar and semantic constraints, enabling deeper exploration of new execution paths. Optimuzz~\cite{Optimuzz} adopts directed greybox fuzzing to generate LLVM IRs targeting specific optimization passes via translation validation.

With the rapid progress of LLMs, many researchers have explored LLM-based compiler testing. Fuzz4All~\cite{Fuzz4All} incorporates LLM-generated prompts into the fuzzing loop to improve test diversity. WhiteFox~\cite{WhiteFox} targets compiler optimizations via a multi-agent 
design leveraging compiler documentation and code examples. LegoFuzz~\cite{ni2025legofuzz} builds programs by composing reusable semantic components. 
Unlike these techniques, \methodName\ further leverages coverage feedback to perform iterative, coverage-driven target selection and prompt refinement, focusing test generation on hard-to-cover compiler code regions.

\subsection{LLM-based Test Generation}\label{sec:8-2}

Beyond compiler testing, LLMs have been widely applied to unit test 
generation~\cite{DBLP:conf/issta/DengXPY023,DBLP:journals/corr/abs-2304-02014,DBLP:conf/kbse/SunYWWJZ23,jiang2024towards} and search-based software testing (SBST)~\cite{CodaMosa,ChatGPT-SBST}. 
For example, TeCo~\cite{TeCo} fine-tunes CodeT5~\cite{CodeT5} to assist developers in completing unit tests, TestPilot~\cite{TestPilot} explores adaptive zero-shot test generation for JavaScript, MuTAP~\cite{MuTAP} employs zero-shot and few-shot prompting and evaluates generated tests via mutation testing, and ChatUniTest~\cite{ChatUniTest} investigates conversation-driven test generation with ChatGPT~\cite{ChatGPT}. CodaMosa~\cite{CodaMosa} generates tests for uncovered methods to guide SBST, ChatGPT-SBST~\cite{ChatGPT-SBST} systematically studies the applicability of ChatGPT in this setting, and CoverUp~\cite{CoverUp} integrates coverage analysis, code context, and iterative feedback to progressively improve statement and branch coverage.

However, applying LLM-guided test generation to compiler testing is more challenging due to complex compilation pipelines and persistent long-tail coverage gaps. To address this, our technique leverages LLM reasoning to analyze uncovered compiler regions and infer triggering requirements (e.g., program patterns and compilation options), guiding test generation toward hard-to-cover code.

\section{Conclusion}\label{sec:conclusion}
To reduce human effort in compiler testing, many test-generation techniques have been proposed; however, they often rely on coarse-grained coverage guidance and fail to effectively explore long-tail code regions. 
In this paper, we propose \methodName, an LLM-based compiler test generation 
technique that targets coverage gaps. \methodName\ iteratively selects 
promising target files via coverage feedback, infers coverage-triggering 
requirements for uncovered regions, and synthesizes prompts to generate 
targeted test programs.
Experiments on GCC and LLVM show that \methodName\ consistently achieves higher source-code coverage than state-of-the-art baselines under the same time budget. Ablation studies further demonstrate the effectiveness of each component of \methodName.
\section{Data Availability}\label{sec:conclusion}
We make the source code and data of \methodName~available on website~\cite{repo}.

\end{document}